# A Tutorial on Explainable Image Classification for Dementia Stages Using Convolutional Neural Network and Gradient-weighted Class Activation Mapping

Kevin Kam Fung Yuen[a,1]

[a]*Department of Computing, The Hong Kong Polytechnic University*
https://orcid.org/0000-0003-1497-2575

**Abstract.** This paper presents a tutorial of an explainable approach using Convolutional Neural Network (CNN) and Gradient-weighted Class Activation Mapping (Grad-CAM) to classify four progressive dementia stages based on open MRI brain images. The detailed implementation steps are demonstrated with an explanation. Whilst the proposed CNN architecture is demonstrated to achieve more than 99% accuracy for the test dataset, the computational procedure of CNN remains a black box. The visualisation based on Grad-CAM is attempted to explain such very high accuracy and may provide useful information for physicians. Future motivation based on this work is discussed.

## 1. Introduction

Dementia is a general term for loss of memory, language, problem-solving and other thinking abilities that are severe enough to interfere with daily life [1]. For example, signs of dementia may include problems with short-term memory, keeping track of a purse or wallet, paying bills, planning and preparing meals, remembering appointments, and travelling out of the neighbourhood. Among different types of dementia, Alzheimer's disease accounts for 60-80% of cases [1].

The dementia's disease progresses in three stages: early, middle and late; or mild, moderate and severe [2] [3]. In the latter stage, the person has more severe brain damage and symptoms. Understanding which dementia stage of a person may help to identify which treatments will be and predict what may occur next. This study presents a tutorial to use a deep learning method to learn the open MRI brain images to predict the dementia stage of a person.

---

[1] Corresponding Author: K.K.F. Yuen, Department of Computing, The Hong Kong Polytechnic University, Hong Kong SAR, China. Email: kevinkf.yuen@gmail.com, kevin.yuen@polyu.edu.hk

Deep convolutional nets have brought about breakthroughs in processing images, video, speech and audio, whereas recurrent nets have shone light on sequential data such as text and speech [4]. Chollet [5] presents a detailed explanation with examples and implementation with Python for deep learning. The first successful practical application of neural nets came in 1989 from Bell Labs, when Yann LeCun combined the earlier ideas of convolutional neural networks and backpropagation, and applied them to the problem of classifying handwritten digits [5]. Lechun et al. [6] introduced the application of convolutional neural networks for classifying handwritten digits.

Regarding the deep learning techniques applied to medical images, Menze et al. [7] report the multimodal brain tumour image segmentation benchmark for twenty tumour segmentation algorithms that were applied to a set of 65 multi-contrast MR scans. Kamal et al. [8] conducted Alzheimer's patient analysis using Image and Gene Expression Data and Explainable-AI to present associated genes. Bae et al. [9] applied a convolutional neural network model based on T1-weighted magnetic resonance imaging to Identify Alzheimer's disease. Qiu et al. [10] presented multimodal deep learning to assess Alzheimer's disease dementia. Marmolejo-Saucedo and Kose [11] applied Grad-Cam Based Explainable Convolutional Neural Network for brain tumour diagnosis.

The contribution of the rest of this article is presented below.

- Data exploration and manipulation of the MRI brain image data are presented in section 3. The image structure, data splitting process and class distribution are presented.
- The explanation and implementation of convolutional neural networks are offered in section 4. A numerical example is shown to explain the meaning of CNN layers used in this paper. Input and output shapes and Parameters are discussed as they are the essential requirements to set a CNN model.
- The gradient-weighted Class Activation Mapping (Grad-CAM), a visual explanation technique for the CNN model, is presented in section 5. The algorithms to plot the heatmap visualisation are presented.
- The simulation environment settings and results are presented in section 5. Training convergence and classification metrics are presented. The Grad-CAM visualisation of CNN predictions for both correct and incorrect instances is presented.
- Conclusion remarks of this study and future motivations based on this study are presented in Section 6.

## 2. Data exploration and manipulation

The Alzheimer dataset is obtained from [12]. The Dataset comprises pre-processed MRI (Magnetic Resonance Imaging) images in *jpg* format of 128 x 128 pixels. A total of 6400 MRI images are divided into four classes presenting four stages of Alzheimer's or dementia: non-dementia (3200 images), very mild dementia (2240 images), mild dementia (896 images), and moderate dementia (64 images). As the moderate dementia group only occupies 1% of the total instances of the dataset and the non-dementia group occupies 50% of the total instances of the dataset, the dataset is imbalanced. Without using some sampling techniques, such as downsampling and upweighting, the simulation shows that the proposed CNN method can handle imbalanced classification properly.

Since the dataset is well pre-processed, augmentation of the image is not needed in this study. The sample instance is visualised in Figure 1.

The dataset is split into three subsets: training, validation and test datasets using the Python *splitfolders* package [13]. The ratio of 80%, 10% and 10% is set for each subset. The seed number is set to 888 for the function for the reproducibility of the same content of the folders. The size of the data subjects after the train-validation-test split is shown in Table 1, i.e., 5119, 639, and 642 different images for each subset. The distribution of the four stages in each subset is shown in Figure 2. The proportions for each class in each subject are similar.

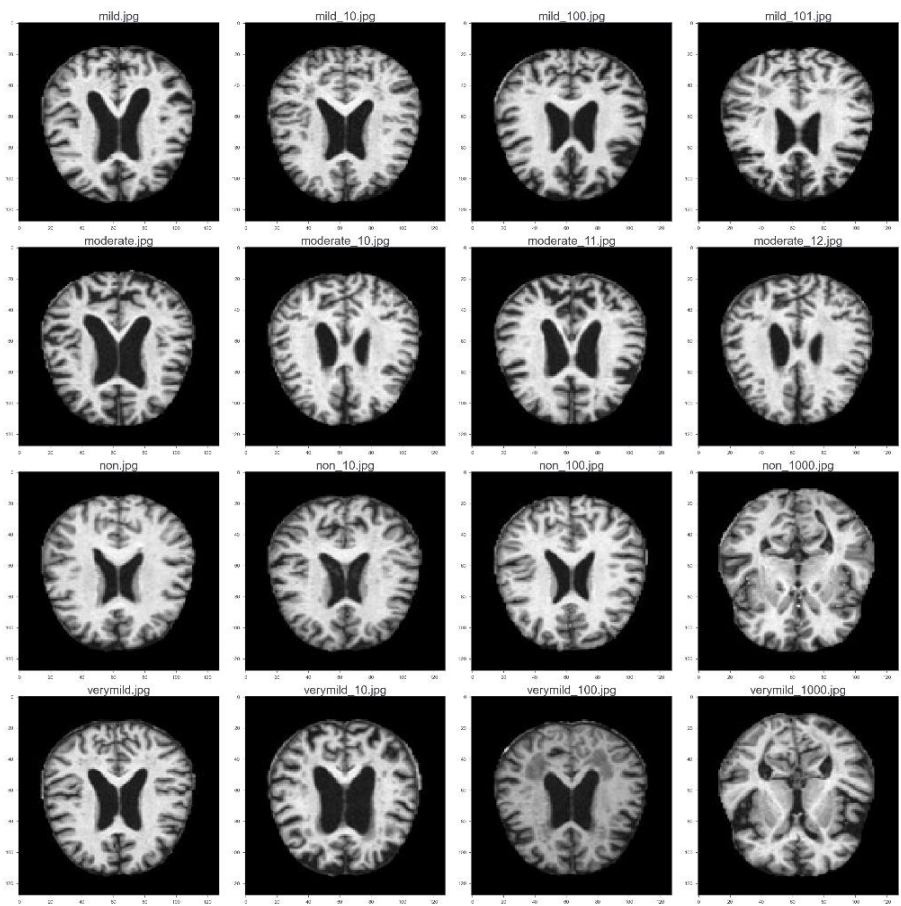

Figure 1. Sample instances of four stage classes

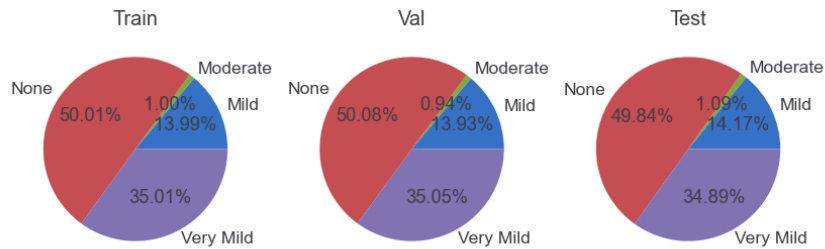

**Figure 2.** Distribution of four stages for each subset

**Table 1.** Size of data subsets after train-validation-test split

| Datasets | None | Very Mild | Mild | moderate | Total |
|---|---|---|---|---|---|
| Training set | 2560 | 1792 | 716 | 51 | 5119 (80%) |
| Validation set | 320 | 224 | 89 | 6 | 639 (10%) |
| Test set | 320 | 224 | 91 | 7 | 642 (10%) |
| Total | 3200 (50%) | 2240 (35%) | 896 (14%) | 64 (1%) | 6400 |

To incorporate image data into the CNN model implemented by TensorFlow and Keras, three *ImageDataGenerator* objects are created to take the images from training, validation and test directories, respectively, by calling the *flow_from_directory* method with rescaling the value between 0 and 1, i.e. dividing a grey image value by 255. For a value in grayscale, 0 means black, and 255 means white. The code statement for an instance of the *ImageDataGenerator* object for training images is shown in Figure 4. Both image height and width are 128 based on the source pre-processed image data.

```
trainIDG = ImageDataGenerator(rescale=1./(255))
trainDatagen = trainIDG.flow_from_directory(
        os.path.join(dataGroups,"train"),
        color_mode="grayscale",
        target_size=(imgHeight, imgWidth),
        class_mode='categorical')
trainDatagen.image_shape

Found 5119 images belonging to 4 classes.
(128, 128, 1)
```

**Figure 4.** Code statement for an *ImageDataGenerator* object of training images

## 3. Convolutional Neural Network

*3.1. Slim CNN Structure*

Deep learning is an Artificial Neural Network (ANN) of layers and each layer consists of many neurons. The word "deep" is not related to a deep understanding of human mind but the depth of an ANN mathematical structure, such as the large scale of the parameters in the network [14]. A convolutional Neural Network is a type of deep learning network that includes at least one convolutional layer(s), pooling layers, and dense (or fully connected) layers. CNN is typically used for image recognition. Keras [15,16] built on top of TensorFlow [17], the Python deep learning API, is utilised to implement the CNN model. A tutorial for the RGB colour images using TensorFlow can be found on the website[2]. CNN settings for this study are mainly based on the grey images due to the nature of the MRI image.

Constructing a CNN model using layers is like playing with Lego to construct buildings using different blocks. Therefore, there are so many types of CNN models, such as AlexNet [18], VGGNet [19], GoogleNet [20], ResNet [21] [22], Xception [23]. To simplify the introduction of notions of CNN, this study proposes a slim CNN structure based on Python code shown in Figure 5, which is much less complicated than the popular ones but achieves more than 99% accuracy for this study. The network consists of one input layer, three 2D convolutional (Conv2D) layers, two 2D max pooling (Maxpooling2D) layers, one Flatten layer, one hidden Dense layer, and one output Dense layer. The implementation code of layers can be found in the documentation[3]. Glossaries for learning machine learning can be found on the Google developer website[4]. The notions of layers and related parameters are briefly explained with examples in the next sections.

```
input = keras.Input((128, 128, 1))
x=layers.Conv2D(128, (3, 3), activation='relu')(input)
x=layers.MaxPooling2D((2, 2))(x)
x=layers.Conv2D(256, (3, 3), activation='relu')(x)
x=layers.MaxPooling2D((2, 2))(x)
x=layers.Conv2D(256, (3, 3), activation='relu', name="lastConv")(x)
x=layers.Flatten()(x)
x=layers.Dense(256, activation='relu')(x)
x=layers.Dense(4,name="output_layer")(x)
model = keras.Model(inputs=input, outputs = x)
```

Figure 5. Code statement of Convolution Neural Network

---

[2] https://www.tensorflow.org/tutorials/images/cnn
[3] https://keras.io/api/layers
[4] https://developers.google.com/machine-learning/glossary

*3.2. Explanation of CNN Layers*

This example demonstrates the basic concept for computation based on the network comprising each consequent layer of input, Conv2D, Maxpooling2D, Flatten, and output. For the input layer, given that the data from 10 to 34 with the shape of (5,5,1), a 5x5 matrix is shown below.

$$\begin{bmatrix} 10 & 11 & 12 & 13 & 14 \\ 15 & 16 & 17 & 18 & 19 \\ 20 & 21 & 22 & 23 & 24 \\ 25 & 26 & 27 & 28 & 29 \\ 30 & 31 & 32 & 33 & 34 \end{bmatrix}$$

Assume that the weights of the (3,3) filter Kernal are from 0.1 to 0.9, as below.

$$\begin{bmatrix} 0.1 & 0.2 & 0.3 \\ 0.4 & 0.5 & 0.6 \\ 0.7 & 0.8 & 0.9 \end{bmatrix}$$

For the output of a Conv2D layer, the convolution matrix with (1,1)-strides is computed as below.

$$\begin{bmatrix} 81.6 & 86.1 & 90.6 \\ 104.1 & 108.6 & 113.1 \\ 126.6 & 131.1 & 135.6 \end{bmatrix}$$

The computation for a convolution matrix is based on the slices of the input matrix with the strides. The slice size is based on the shape of the filter kernel. For example, the entry of 86.1 of the convolutional matrices is calculated by the form below.

$$sum\left( \begin{bmatrix} 11 & 12 & 13 \\ 16 & 17 & 18 \\ 21 & 22 & 23 \end{bmatrix} \circ \begin{bmatrix} 0.1 & 0.2 & 0.3 \\ 0.4 & 0.5 & 0.6 \\ 0.7 & 0.8 & 0.9 \end{bmatrix} \right) = sum\left( \begin{bmatrix} 1.1 & 2.4 & 3.9 \\ 6.4 & 8.5 & 10.8 \\ 14.7 & 17.6 & 20.7 \end{bmatrix} \right) = 86.1$$

Regarding the Maxpooling2D layer, with a pool size of (2, 2) and strides of (1, 1), the max pool for the output of the convolution matrix is computed below.

$$\begin{bmatrix} 108.6 & 113.1 \\ 131.1 & 135.6 \end{bmatrix}$$

Strides of (1, 1) mean moving the slice (filter) one step (column) to the left if not reaching the end of the columns and moving the slice one step (row) down to the start of the column if reaching the end of the columns. For the entry of 135.6, the calculation is shown below.

$$max\left( \begin{bmatrix} 108.6 & 113.1 \\ 131.1 & 135.6 \end{bmatrix} \right) = 135.6$$

For the Flatten layer, the data form is transformed from 2D matrix to a 1D vector,
$$[108.6, 113.1, 131.1, 135.6].$$

Regarding the output dense layer, supposed that there are two classes, therefore, two nodes. Each node has different weights such that the Flatten layer to Node A is [0.3,0.2,0.4,0.1] and the Flatten layer to Node B is [0.2,0.2,0.5,0.1]. Therefore, the output value after the weighted sum of the two vectors respectively is [121.2, 123.45]. Tentatively, the input is predicted to be Class B.

The weights in the convolutional layers and dense layers are continuously updated until the highest accuracy and lowest loss reach. The *relu* activation function converts the negative values to 0, and this example does not have any negative values. The weights for the CNN network are used to predict the new instance. The calculation of the CNN parameters is discussed in the next section.

*3.3. Input/Output Shapes and Parameters*

The name for deep learning may be regarded as another branding name for artificial neural network (ANN) due to the large scale of parameters of the more complicated structure to be explored, which was very hard to find the optimal values based on the legacy machine, but much earlier to do with today's hardware technology. Table 2 shows the summary of the Convolution Neural Network Structure shown in Figure 5. There are 52,268,036 total parameters, which are all trainable. The word "Trainable" may mean finding a good-enough (ideally best) value for a parameter, and thus CNN is a heuristic algorithm. Input, maxPooling2D, and flatten layers do not have any tunning parameters for operations. The output of the previous layer is the input of the current layer.

**Table 2.** Summary of Proposed Convolution Neural Network Structure

| Layer(Code) | Output Shape | Param # |
| --- | --- | --- |
| input = keras.Input((128, 128, 1)) | (None, 128, 128, 1) | 0 |
| x=layers.Conv2D(128, (3, 3), activation='relu')(input) | (None, 126, 126, 128) | 1,280 |
| x=layers.MaxPooling2D((2, 2))(x) | (None, 63, 63, 128) | 0 |
| x=layers.Conv2D(256, (3, 3), activation='relu')(x) | (None, 61, 61, 256) | 295,168 |
| x=layers.MaxPooling2D((2, 2))(x) | (None, 30, 30, 256) | 0 |
| x=layers.Conv2D(256, (3, 3), activation='relu', name="lastConv")(x) | (None, 28, 28, 256) | 590,080 |
| x=layers.Flatten()(x) | (None, 200704) | 0 |
| x=layers.Dense(256, activation='relu')(x) | (None, 256) | 51,380,480 |
| x=layers.Dense(4,name="output_layer")(x) | (None, 4) | 1,028 |

Total (trainable) params: 52,268,036 (199.39 MB)

For the input layer, the input and output have the shape of (batch, height, width, and depth). "None" elements in the shape represent dimensions where the shape is not known and may vary. The batch size is not defined and is subject to the variable. For a grey image, the channel value is 1. For an RGB colour image, the channel is 3. The output shape is the same as the input shape of (None, 128, 128, 1), which is an image with a height of 128, width of 128, and depth (or channel) of 1.

For the *Conv2d* parameters, the output size is

$$\text{Conv2dSize} = \frac{inputSize - filterSize + 2 \times paddingSize}{strideSize} + 1 \qquad (1)$$

Since stride size is one and padding size is valid, i.e., zero, by default, the default size of *Conv2D* is as below.

$$\text{conv2dSize} = inputSize - filterSize + 1 \qquad (2)$$

For the first *Conv2D* layer, the output shape of (height and width) is

$$(128,128) - (3,3) + 1 = (126,126)$$

Since the filter number is set to 128 filters, the output shape is (None, 126, 126, 128). The total number of parameters is calculated in the form below.

$$\text{Conv2d parameters} = \text{Filter parameters} + \text{Bias parameters} \qquad (3)$$

Thus, the number of parameters is

$$(3 \times 3 \times 1) \times 128 + 128 = 1280$$

For each kernel filter, $(3 \times 3 \times 1)$ weights must be optimised, and there are 128 filters and 128 bias parameters.

For the first MaxPooling2D layer, the input to max pooling is the output of conv2d when the input shape is larger than or equal to the pool size. The output of height and width is calculated in the form below.

$$\text{MaxPooling2dSize} = \left\lfloor \frac{inputShape - poolSize}{strides} \right\rfloor + 1 \qquad (4)$$

Strides are set to *none* by default, which means the same as the pool size. The size of height and width is

$$\left\lfloor \frac{(126,126) - (2,2)}{(2,2)} \right\rfloor + 1 = (63,63).$$

For the second *Conv2D* layer, the input shape of height and width is

$$(63,63) - (3,3) + 1 = (61,61).$$

As the depth is 128 in previous layers and there are 256 filters, the number of parameters is

$$(3 \times 3 \times 128) \times 256 + 256 = 295{,}168.$$

For the second *MaxPooling2D* layer, the input shape of height and width is

$$\left\lfloor \frac{(61,61) - (2,2)}{(2,2)} \right\rfloor + 1 = (30,30)$$

For the third *Conv2D* layer, the input shape of height and width is

$$(30,30) - (3,3) + 1 = (28,28).$$

As both depth and filters are 256, the number of parameters is

$$(3 \times 3 \times 256) \times 256 + 256 = 590{,}080.$$

For the *Flatten* layer, the input vector shape is

$$28 \times 28 \times 256 = 200{,}704.$$

For the hidden dense layer, the number of neurons is set to 256. The number of biases is 256. The number of parameters is

$$200{,}704 \times 256 + 256 = 51{,}380{,}480.$$

For the output dense layer, the number of neurons is set to 4. The number of biases is 4. The number of parameters is

$$4 \times 256 + 4 = 1028.$$

Finally, the total number of trainable parameters is

$$1280 + 295168 + 590080 + 51380480 + 1028 = 52{,}268{,}036.$$

As the computational workload to optimise more than 52 million parameters is very high for this case, that's the name for deep learning instead of ANN. Understanding the calculations above can help us configure the right hyper-parameters for settings.

**4. Gradient-weighted Class Activation Mapping**

The gradient-weighted Class Activation Mapping (Grad-CAM) is a 'visual explanations' technique for decisions from a large class of CNN-based models, making them more transparent and explainable [24,25]. Grad-CAM applies the gradients of a target concept to the final convolutional layer to produce a coarse localisation map to highlight the important regions in the image for predicting the concept.

Implementation of Grad-CAM for the proposed CNN model is slightly revised from the demo code [5], which is also shown in the Keras web[5]. Further modification of the algorithm is needed for the proposed CNN based on the structure mentioned in Sect. 4 and the MRI image is in grey channel instead of RGB channels. The Grad-CAM visualisation is summarised in Algorithms 1-3. In general, Algorithm 3 calls Algorithm 2 calling Algorithm 1. Figures 10 and 11 are the examples of the results from using Algorithm 3. The colourmap is produced by matplotlib[6]. Figure 6 shows the Grey and jet heatmap gradient colourmaps used in this paper.

| Algorithm 1: Grad-CAM Heatmap Algorithm (makeGradcamHeatmap) |
|---|
| Input: Image array, CNN Model, name of last convolution layer |
| 1. create a Keras model that maps the input image to the activations of the last conv layer as well as the output predictions |
| 1. compute the gradient of the top predicted class for the input image array with respect to the activations of the last conv layer |
| 2. compute the gradient of the output neuron (top predicted or chosen) regarding the output feature map of the last conv layer |
| 3. compute a vector of each entry of the mean intensity of the gradient over a specific feature map channel |
| 4. multiply each channel in the feature map array by "how important this channel is" regarding the top predicted class, then sum all the channels to obtain the heatmap class activation |
| 5. Normalize the heatmap between 0 and 1 |
| Return: heatmap array |

---

[5] https://keras.io/examples/vision/grad_cam
[6] https://matplotlib.org/stable/users/explain/colors/colormaps.html

| Algorithm 2: Grad-CAM Display Algorithm (GradcamDisplay) |
|---|
| Input: Image, CNN Model, name of last convolution layer |
| 1. Convert the image file into array data
2. Preprocessing the image array by inserting a new dimension
3. Create a Gra-CAM heatmap by calling Algorithm 1.
4. Rescale heatmap to a range between 0 and 255
5. Set a colour map scheme such as Jet.
6. Convert the scaled heatmap value
7. Create a colormap-heatmap image with RGB colourised heatmap, e.g., an image with RGB colourised heatmap
8. Superimpose the heatmap on the original image by
   superimposed_img = colormap_heatmap * alpha + img |
| Return: raw grey heatmap in step 3, colormap-heatmap image, superimposed image |

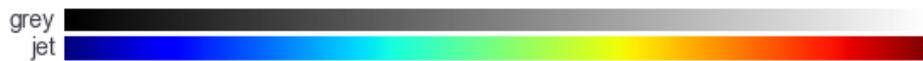

Figure 6. Grey and Jet Heatmap Gradient Colormaps

| Algorithm 3: Grad-CAM Explainability Plot Algorithm (PlotCases) |
|---|
| Input: cases of images, m =4 |
| 1. If the case is supplied, the case is used, and the *m* value is updated. Otherwise, randomly select *m* cases.
2. Create a plot matrix with 3 columns and m rows
3. For each row, plot the original, grey heatmap, jet heatmap and prediction images for each column by calling algorithm 2. |
| Return: an image including original, grey heatmap, jet heatmap and prediction sub-images selected or randomised. |

## 5. Simulation and Discussions

The code is developed in Python Jupyter Notebook. The source code of this study is available in the author's GitHub[7]. The program was executed on the machine Lenovo Legion Pro 5 16IRX8 with i9-13900HX CPU and 32 GB RAM. As the details are summarised in Table 1, 5119 images (80% of the total) are used to train the proposed CNN model, 639 images (10% of the total) are used to validate the model, and 642 images (10% of the total) are used to test the model performance. The distribution for each class is imbalanced.

---

[7] https://github.com/kkfyuen/ (to be uploaded)

```
model.compile(optimizer='adam',
              loss=tf.keras.losses.CategoricalCrossentropy(from_logits=True),
              metrics=['accuracy'])
modelCheckpointer = ModelCheckpoint(filepath="weights1.keras", verbose=1,
                                                                save_best_only=False)
tensorboard_callback = keras.callbacks.TensorBoard(log_dir="logs/fit")
history = model.fit(trainDatagen,
                    epochs=50,
                    validation_data=valDatagen,
                    callbacks=[modelCheckpointer, tensorboard_callback])
```

Figure 7. Code for training CNN

Figure 7 shows the code for training the CNN model. Although the tensor board and model checkpoint are optional for fitting the model, the tensor board is used to view the training metric performance, whilst the model checkpoint is used to save the values of the parameters, and thus, the result is reproducible for the prediction. 50 epochs are set for training. The validation data is used to validate during the training process.

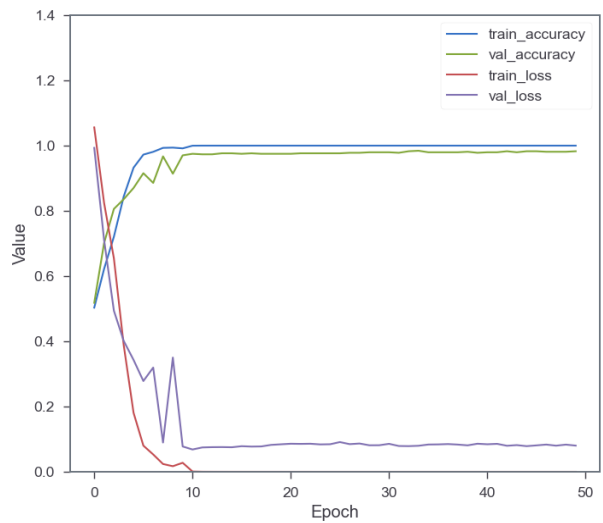

Figure 8. training and validation convergences based on accuracy and loss metrics

Figure 8 exhibits the training and validation convergences based on accuracy and loss metrics. The entire training process took 6531 seconds (1.81 hours). The accuracy and loss start to converge after 10 epochs. The training accuracy is 100%, whilst the validation accuracy is 98.28%. The training loss is 0%, and the validation loss is 8%. Based on the dataset, the model has an excellent (very close to perfect) and very stable fit.

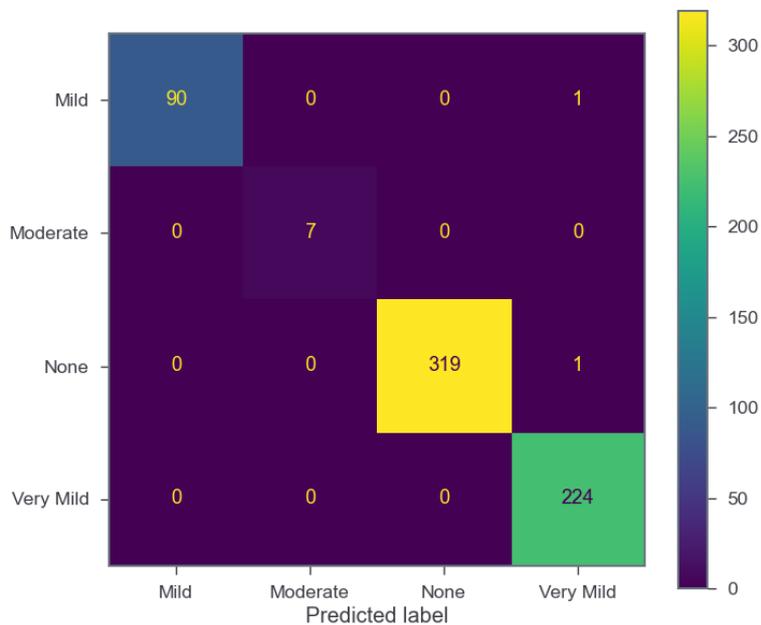

Figure 9. Confusion Matrix of classification for test dataset

Figure 9 shows the confusion matrix of classification for the test dataset for the testing dataset. Surprisingly, only two instances are misclassified. The accuracy is 640/642=99.69%. Based on the confusion matrix, Table 3 shows the classification metrics report, including precision, recall, and F1-score using the *ConfusionMatrixDisplay* function in *Scikit-learn* package[8]. The values of precision, recall and F1-score are very close to 100%. The proposed CNN for this dataset is a closely perfect solution.

**Table 3.** Classification metrics report (*round to two decimal digits)

| Dementia stage | Precision | Recall | F1-score | Support |
|---|---|---|---|---|
| None | 1 | 1 | 1 | 319 |
| Very Mild | 1 | 0.99 | 1 | 226 |
| Mild | 0.99 | 1 | 0.99 | 90 |
| Moderate | 1 | 1 | 1 | 7 |
| accuracy | | | 1 | 642 |
| macro avg | 1 | 1 | 1 | 642 |
| weighted avg | 1 | 1 | 1 | 642 |

---

[8] https://scikit-learn.org/stable/modules/generated/sklearn.metrics.ConfusionMatrixDisplay.html

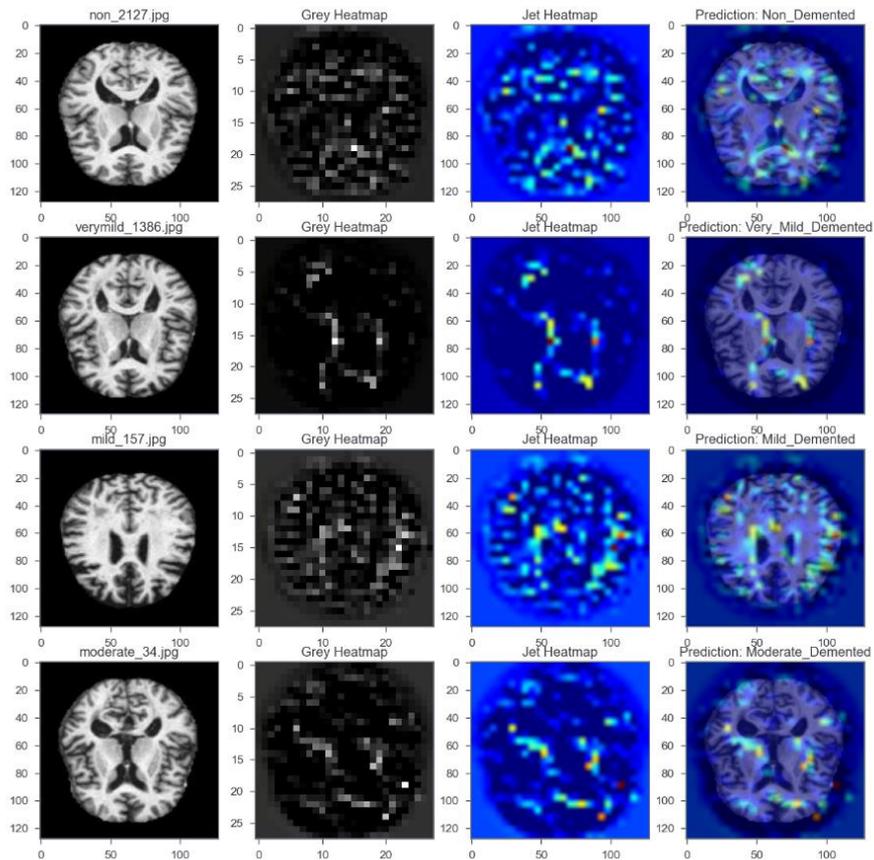

Figure 10. Grad-CAM images of correct classification

Although the proposed CNN has an almost perfect prediction result (99.69%), the algorithm is unknown to the user. Grad-CAM may be used to explain the algorithm reasoning results by visualising the important area in the image with a heatmap. Figure 10, produced by Algorithm 3, shows an instance for each stage class. the original image is first displayed, then a grey heatmap is produced based on the original image and the last convolution layer, a jet heatmap is further displayed based on the grey heatmap, and finally, the combined image is based on the jet heatmap, and the original photo is displayed. The combined image attempts to show the important pixels for the prediction results.

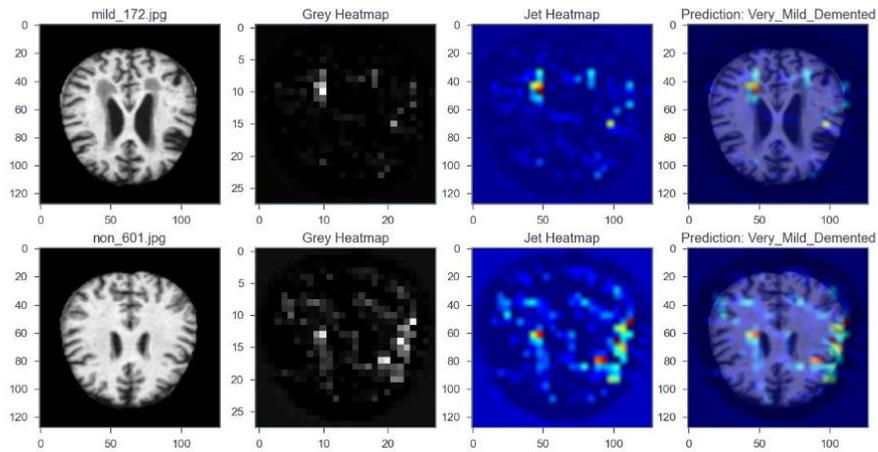

Figure 11. Grad-CAM images of incorrect classification

Two cases, *mild_172.jpg* and *non_601.jpg,* are misclassified as very mild and is misclassified. Figure 11 shows the heatmap pixels for the prediction results. The Gra-CAM method may exhibit the visualisation to explain why the result is not correct based on the highlighted area. Although the CNN method is valuable and the Gra-CAM visualisation is promising, the heatmap results require further justification from the medical partitioners, which is suggested for future study.

## 6. Conclusion

This study introduces a slim structure of CNN comprising only nine layers, which produces more than 99% accuracy for dementia stage prediction using the open dataset in Kaggle. The layers and parameter settings are explained with examples. As the explainability of an almost perfect solution is unclear in CNN, this study presents the use of Grad-CAM to visualise the explainability of the proposed CNN. Whilst the focus of this study is mainly based on deep learning classification, the major limitations are that the patterns and implications of medical findings based on visualisation need to be further explored and evaluated, which may be a motivation for future medical research. This study presents a tutorial of the practical application with detail implementation and explanation. Future works may include the exploration and comparisons of different CNN structures with different explainable visualisation methods.

**Code availability**
The work for the study will be available on the author's GitHub, https://github.com/kkfyuen/.